\documentclass{ws-mpla}
\newcommand{\be}{\begin{equation}}
\newcommand{\ee}{\end{equation}}
\def\eg{{\it e.g.~}}

\def\bone{B^{(1)}}

\begin{document}

\markboth{Bertone and Merritt}
{Dark Matter Dynamics and Indirect Detection}

%
\catchline{}{}{}{}{}
%

\title{DARK MATTER DYNAMICS AND INDIRECT DETECTION}

\author{GIANFRANCO BERTONE}

\address{NASA/Fermilab Particle Astrophysics Center,\\
Batavia, IL 60510, USA\\
bertone@fnal.gov}

\author{DAVID MERRITT}

\address{Department of Physics, Rochester Institute of Technology,\\
54 Lomb Memorial Drive, Rochester, NY 14623, USA\\
merritt@astro.rit.edu}

\maketitle


\begin{abstract}
Non-baryonic, or ``dark'', matter is believed to be a 
major component of the total mass budget of the universe.
We review the candidates for particle dark matter
and discuss the prospects for direct detection
(via interaction of dark matter particles with laboratory
detectors) and indirect detection (via observations
of the products of dark matter self-annihilations),
focusing in particular on the Galactic center, which is  
among the most promising targets for indirect detection studies.
The gravitational potential at the Galactic center
is dominated by stars and by the supermassive black
hole, and the dark matter distribution is expected to
evolve on sub-parsec scales due to interaction with
these components.
We discuss the dominant interaction mechanisms
and show how they can be used to rule out certain
extreme models for the dark matter distribution,
thus increasing the information that can 
be gleaned from indirect detection searches.

\end{abstract}


\section{Dark Matter Candidates and Constraints}

If one compares the total amount of matter inferred from 
cosmic microwave background (CMB) experiments\cite{Spergel:2003cb}
\be
\Omega_M h^2=0.135^{+0.008}_{-0.009}.
\label{omh2}
\ee
(here $\Omega$ denotes the mean density as a fraction of the
critical, or closure, density, and $h$ is the Hubble constant
in units of $100$ km s$^{-1}$ Mpc$^{-1}$)
with the amount of baryons allowed by the CMB
\be
\Omega_b h^2= 0.0224\pm 0.0009 \,
\ee
(also consistent with the Big Bang nucleosynthesis constraint
$0.018 < \Omega_b h^2 < 0.023$, see \eg Ref.~\refcite{Olive:2003iq}), one
is left with a component of the Universe that is composed of
matter distinct from ordinary baryonic matter. 
There is no shortage of dark matter candidates, most of them arising
in theories beyond the standard model of particle physics. 
In the framework of the standard model, neutrinos have been 
proposed as dark matter candidates, but the analysis of CMB
anisotropies, combined with large-scale structure data,
suggest $\Omega_{\nu} h^2 < 0.0067$ (95\% confidence limit),
which implies that neutrinos are a sub-dominant component
of non-baryonic dark matter. 

Rather than compile a complete list of all possible dark matter 
candidates (often referred to as {\it WIMPs}: ``Weakly Interacting 
Massive Particles'') we discuss here the candidates that have received
the widest attention in the recent literature. For an extensive 
review of particle dark matter candidates see e.g. 
Refs.~\refcite{Bertone:2004pz} and~\refcite{Bergstrom:2000pn}.

The {\it neutralino} in models of R-parity-conserving supersymmetry is 
by far the most widely studied dark matter candidate. 
In the ``minimal supersymmetric standard model'' (MSSM), 
the superpartners 
of the $B$ and $W_3$ gauge bosons (or the photon and $Z$, equivalently) and 
the neutral Higgs bosons, $H_1^0$ and $H_2^0$, are called binos ($\tilde{B}$), 
winos ($\tilde{W_3}$), and higgsinos ($\tilde{H_1^0}$ and $\tilde{H_2^0}$), 
respectively. 
These states mix into four Majorana fermionic mass eigenstates, 
called neutralinos. 
The lightest of the four neutralinos, commonly referred to as 
``the'' neutralino, is in most supersymmetric scenarios the lightest 
supersymmetric particle. 
The neutralino is a perfect dark matter candidate, 
with mass and interaction cross sections that, in fine-tuned regions of the
supersymmetric parameter space, can correctly reproduce the
dark matter relic density while still being consistent with constraints
on the WIMP mass and searches in accelerators as well as direct and indirect
detection experiments.

Another interesting candidate arises in theories with universal
extra dimensions (UED),\cite{Appelquist:2000nn}
i.e. extra-dimension scenarios in which all fields 
are allowed to propagate in the bulk. 
Upon compactification of the extra dimensions, all of the fields propagating 
in the bulk have their momentum quantized in units of $p^2 \sim 1/R^2$, 
where $R$ is the compactification radius, appearing as a tower 
of states with masses $m_n= n/R$, where $n$ labels the mode number. 
Each of these new states contains the same quantum numbers
(charge, color, etc.)
The {\it lightest Kaluza-Klein particle} (LKP) in the framework of UED,
which is likely to be associated with the first KK 
excitation of the photon, or more precisely the first KK excitation 
of the hypercharge gauge boson,\cite{Cheng:2002iz} provides a viable
dark matter candidate. 
We will refer to this state as $\bone$. 
The  calculation of the $\bone$ relic density was
performed by Servant and Tait,\cite{Servant:2002aq} 
who found that if the LKP is to account for the observed quantity of 
dark matter, its mass (which is inversely proportional to the 
compactification radius $R$) should lie in the range 400 to 1200 GeV, 
well above any current experimental constraint.

Dark matter candidates are commonly believed to have masses in the range
1 GeV -- 100 TeV. 
The lower value is the so-called Lee-Weinberg 
limit,\cite{leeweinberg,Hut:1977zn} 
while the upper value comes from the so-called unitarity 
bound.\cite{weinberg,Griest:1989wd} It is sometimes stated in the 
literature that the upper bound is of order 340 TeV. 
However, this comes from the now-obsolete requirement that
$\Omega_M h^2 \lesssim 1$, adopted in 1989 by Griest and 
Kamionkowski.\cite{Griest:1989wd}.
If one instead adopts the modern estimate in Eq. (\ref{omh2}), 
the upper limit on the mass of the dark matter particle is 
$m \lesssim 120$TeV.
 
These limits are, however, model-dependent.
Light scalar particles with masses below 1 GeV are viable dark matter 
candidates,\cite{mevproposed1,mevproposed2} 
(referred to as {\it light dark matter} or alternatively 
{\it MeV dark matter}), 
since the Lee-Weinberg limit strictly applies only to fermionic particles
with standard model interactions. 
If MeV dark matter is to explain 
the 511 keV emission from the Galactic bulge observed by INTEGRAL,
\cite{Jean:2003ci,Knodlseder:2003sv,Weidenspointner:2004my} then
a comparison of the inverse Bremsstrahlung emission (associated with 
dark matter annihilations into electron-positron pairs) 
with the diffuse Galactic background constrains the mass of the 
light dark matter particle to be smaller than 
$\sim 20$ MeV.\cite{Beacom:2004pe} 
It has been argued that hints of this scenario may also have 
been discovered in 
particle physics experiments~\cite{Boehm:2004gt} and could have interesting 
implications for neutrino physics.\cite{Boehm:2004uq}

The unitarity bound can also be violated
if, e.g., dark matter particles were not in thermal equilibrium in the 
early Universe, but were instead produced via alternative mechanisms,
such as gravitational production.\cite{Chung:2001cb,Chang:1996vw}

\section{Halo Models}

The probability of direct detection is proportional to the 
dark matter density in the solar neighborhood, and most discussions of
indirect detection have also focussed on the Milky Way; in both cases, 
a model of the Milky Way's dark matter halo is crucial for predicting
and interpreting event rates.
Halo density profiles are usually derived from $N$-body simulations 
of gravitational clustering, which predict a characteristic form
for $\rho(r)$, and on dynamical constraints like the
Sun's orbital velocity, which provide a normalization.
(For reviews, see Refs. \refcite{Primack:03} -- \refcite{Navarro:04}.)
The most recent and highest-quality 
$N$-body simulations\cite{Navarro:04b,Reed:05} 
suggest a universal dark-matter density profile of the form
\begin{equation}
\rho(r) \approx \rho_0 \exp\left[-\left(r/r_0\right)^{1/n}\right]
\label{eq:nfw}
\end{equation}
with $n\approx 5$;\cite{Merritt:05} the density normalization at the Solar
circle implied by the Galactic rotation curve is 
$\rho\approx 0.3$ GeV cm$^{-3}\approx 8\times 10^{-3}M_\odot{\rm pc}^{-3}$,
with an uncertainty of a factor $\sim 2$.
Henceforth we refer to Eq. (\ref{eq:nfw}) with 
$\rho(R_\odot)=0.3 {\rm GeV} {\rm cm}^{-3}$ as the {\it standard halo model}
(SHM).
(Eq. \ref{eq:nfw} replaces the more approximate 
NFW model\cite{NFW:96} in which the density is a single power-law,
$\rho\sim r^{-1}$, inward of the Solar radius $R_\odot$.)
Unfortunately the $N$-body simulations from  which these
halo models are derived have resolutions that are 
measured in hundreds of parsecs at best, 
whereas the signal from dark matter annihilations depends
critically on the mass profile in the inner few parsecs of
the Galaxy.
Furthermore the $N$-body simulations typically ignore the
influence of the baryons (stars, gas, etc.) even though these
are known to dominate the gravitational potential of the inner Galaxy.

One simple, and probably highly idealized, 
way to account for the effect of the baryons on the dark matter
is via {\it adiabatic contraction}  models,
which posit that the baryons contracted quasi-statically and
symmetrically within the pre-existing dark matter halo, 
pulling in the dark matter and increasing its density.\cite{Blumenthal:86}
When applied to a dark matter halo with the density law
(\ref{eq:nfw}), the result is a halo profile with $\rho\sim r^{-\gamma_c}$,
$\gamma_c\approx 1.5$ inward of the Solar circle and an
increased density at $R_\odot$.\cite{Edsjo:04,Prada:04,Gnedin:04a}
Alternatively, strong departures from spherical symmetry -- 
for instance, during the mergers that created the Galactic bulge
-- might have resulted in {\it lower} dark matter 
densities in the inner tens of parsecs.\cite{Ullio:01,Merritt:02,MM:02}
However we argue below that the steeply-rising {\it stellar} density
near the Galactic center makes such models unlikely, at
least in the case of the Milky Way.

The annihilation signal from a region of volume $V$ 
that includes the Galactic center is
proportional to $\langle\rho^2\rangle V$.
For any $\rho\propto r^{-\gamma}$ with $\gamma\ge 3/2$, the
small-radius dependence implies a divergent flux;
hence the strength of an annihilation signal is 
crucially dependent on the dark matter distribution
on very small scales, where neither the simulations
nor the dynamical data provide useful constraints.
In the following section we discuss physical arguments
that can be used to constrain the dark matter distribution
on these small scales, and in the final section we
discuss the implications for indirect detection.

\section{Dark-Matter Dynamics on Sub-Parsec Scales}

At distances $r\lesssim 1$ pc from the Galactic center,
the gravitational potential is dominated by stars 
and by the supermassive black hole (SBH).
The stars are observed to have a density\cite{Genzel:03}
\begin{equation}
\rho_\star(r)\approx 3.2\times 10^5M_\odot {\rm pc}^{-3}(r/1{\rm pc})^{-1.4},\ \ \ \ r\lesssim 5 {\rm pc}.
\label{eq:nucdens}
\end{equation}
Estimates of the SBH's mass range from
$2-4\times 10^6M_\odot$,\cite{Chakrabarty:01,Ghez:03,Schoedel:03};
the most recent estimates, based on the orbits of single
stars, suggest $\sim 4\times 10^6M_\odot$,
but for consistency with most pre-2005 papers on dark
matter in the Galactic center, 
we here adopt $M_{bh}=3.0\times 10^6M_\odot$.
The gravitational influence radius $r_h$ of  the SBH
(the radius containing a mass in stars equal
to twice the SBH's mass) is $r_h\approx 2$ pc.
Within $r_h$, the dark matter distribution is likely to
have been strongly affected by whatever processes resulted
in the formation of the SBH and the nuclear star
cluster, and by any subsequent interactions between dark
matter and stars.

Fig. 1 shows a number of possible models for the dark
matter density on sub-parsec scales.
The standard halo  model (SHM) discussed above,
normalized to
$\rho=0.3$ GeV cm$^{-3}$ at the Solar circle,
predicts $\rho(r_h)\approx 10^{1.5}M_\odot {\rm pc}^{-3}$.
Allowing for adiabatic contraction of the dark matter
by the baryons (\S 2) increases this by a factor of $\sim 10^2$,
and the inner density slope $\gamma\equiv -d\log\rho/d\log r$ in this 
contracted model is
$\sim 1.5$ or steeper, implying a divergent annihilation flux.
But even steeper density profiles are possible.
If the SBH grew in the simplest possible way --
via slow, symmetrical infall of gas -- the density
of matter around it would also grow, in the same way
that contracting baryons steepen the overall Galactic 
dark matter profile.\cite{Peebles:72,Young:80}
This scenario predicts a final density (of stars or dark matter)
near the SBH of
\begin{equation}
\rho(r) \approx \rho(r_{sp})\left({r\over r_{sp}}\right)^{-\gamma_{sp}},\ \ \ \ \gamma_{sp}=2+{1\over 4-\gamma_c}
\end{equation}
with $\gamma_c$ the dark matter density slope in the 
pre-existing ``cusp,'' i.e. the region $r\gtrsim r_h$.\cite{Gondolo:99}
Such a {\it spike} in the dark matter density would extend inward from 
$r_{sp}\approx 0.2r_h$,\cite{Merritt:04a}
and for $\gamma_c\gtrsim 1$, $\gamma_{sp}\gtrsim 2.3$ (Fig. 1).
Now, the {\it stellar} density profile would respond in a similar way 
to growth of the SBH, and $\rho_\star(r)$ is known to
be shallower than $\rho_*\sim r^{-2.3}$ (Eq. \ref{eq:nucdens}). 
But this is probably a result of
dynamical evolution that occurred
after the SBH was in place (see below).

At even smaller radii,
a limit to the dark matter density is set by self-annihilations:
$\rho\lesssim\rho_a\equiv m/\langle\sigma v\rangle t$, 
with $t\approx 10^{10}$ yr the time since formation of the 
spike.\cite{Berezinsky:94}
A strict inner cut-off to the dark matter density is set by the
black hole's event horizon, $r_{Sch}=2GM_{bh}/c^2\approx 3\times 10^{-7}$ pc
for a non-rotating hole, although for reasonable values of
$m$ and $\langle\sigma v\rangle$, the density is limited by
self-annihilations well outside of $r_{Sch}$ (Fig. 1).

\begin{figure}[th]
\centerline{\psfig{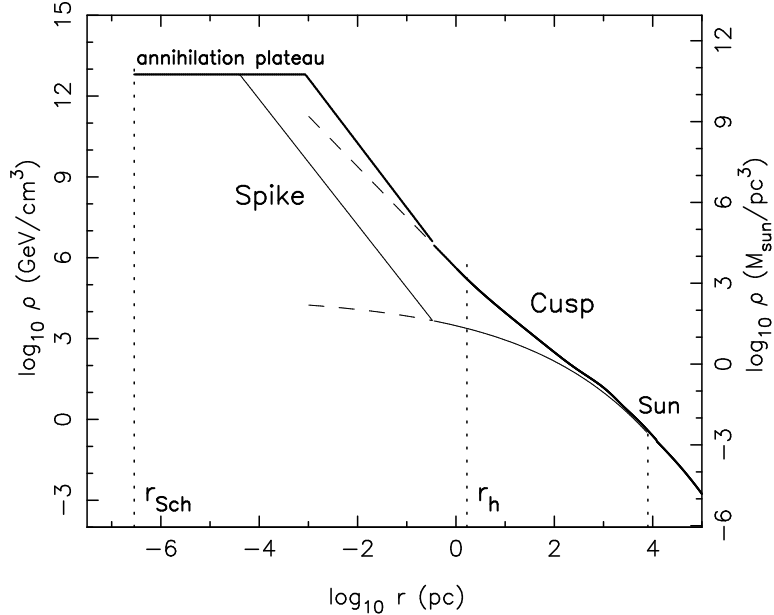}}
\vspace*{8pt}
\caption{Possible models for the dark matter distribution in the
Galaxy.
The thin curve shows the standard halo model (SHM),
and the thick curve is the same model after 
``adiabatic compression'' by the Galactic baryons 
(stars and gas).
Both curves are normalized to a dark matter density of
$0.3$ GeV cm$^{-3}$ at the Solar circle.
Curves labelled ``spike'' show the increase in density that
would  result from growth of the Galaxy's (SBH) at
a fixed location.
The annihilation plateau, $\rho=\rho_a=m/\langle\sigma v\rangle t$, 
was computed assuming $m=200$ GeV,
$\langle\sigma v\rangle=10^{-28}$ cm$^3$ s$^{-1}$ and $t=10^{10}$ yr.
Dashed vertical lines indicate the SBH's Schwarzschild
radius ($r_{Sch}\approx 2.9\times 10^{-7}$ pc,
assuming a mass of $3.0\times 10^6M_\odot$ and zero rotation),
the SBH's gravitational influence radius ($r_h\approx 1.7$ pc),
and the radius of the Solar circle ($R_\odot\approx 8.0$ kpc).
Effects of the dynamical processes discussed in this article
(scattering of dark matter off stars, loss of dark matter
into  the SBH, etc.) are excluded from this plot;
these processes would generally act to decrease the dark matter
density below what is shown here, particularly in the models
with a ``spike.''}
\end{figure}

We will discuss in the next section the prospects of observing DM annihilation
radiation from the Galactic center. The dependence of the annihilation signal 
on the DM profile is usually contained in the factor $J$ defined as
\begin{equation}
\overline{J}_{\Delta\Omega} = K\Delta\Omega^{-1}\int_{\Delta\Omega} d\psi\int_\psi \rho^2 dl
\label{eq:jdo}
\end{equation}
with $\Delta\Omega$ the angular acceptance of the detector,
$dl$ a distance increment along the line of sight, and
$\psi$ the angle between the line of sight and the Galactic
center; the normalizing factor $K$ is normally set to
$K^{-1}=(8.5{\rm kpc})(0.3{\rm GeV}/{\rm cm}^3)^2$.
Henceforth we denote by $\overline{J}_3$ and $\overline{J}_5$
the $\overline{J}$ values corresponding to 
$\Delta\Omega=10^{-3}$ sr and $10^{-5}$ sr respectively;
the former is the approximate angular acceptance
of EGRET\cite{EGRET} while the latter corresponds approximately to
atmospheric Cerenkov telescopes (ACTs) like VERITAS\cite{VERITAS},
CANGAROO\cite{CANGAROO} and HESS\cite{HESS}
and to the proposed satellite observatory GLAST.\cite{GLAST}

In dark matter models with an inner spike it is possible to solve analytically 
the integral of $\rho^2$, and the resulting expression for $J$ is
\begin{eqnarray}
\overline{J}_{\Delta\Omega} &\approx& {4\pi\over 3} {J_0\over \Delta\Omega} 
{\rho_a^2r_a^3 \over R_\odot^2}
\left\{ 1+{3\over 2\gamma_{sp}-3}\left[1-\left({r_a\over r_{sp}}
\right)^{2\gamma_{sp}-3} \right]\right\} \nonumber \\
&\approx& {4\pi\over 3} {J_0\over \Delta\Omega} 
{\rho_a^2r_a^3 \over R_\odot^2}
{1\over 1-3/(2\gamma_{sp})} \nonumber \\
&\approx& {10\over \Delta\Omega} 
\left({\rho_a\over\rho_\odot}\right)^2
\left({r_a\over R_\odot}\right)^3
\label{eq:japprox}
\end{eqnarray}
with $r_a$ the outer radius of the region where the density is
limited by self annihilations (Fig. 1); 
these expessions assume that
$\Delta\Omega\gg (r_{sp}/R_\odot)^2$ and
the latter two expressions assume $r_a\ll r_{sp}$.
A feeling for the range of plausible $\overline J$ values can be had
by computing the $\rho_a$ and $r_a$ values corresponding to a set of ten,
minimal supergravity benchmark models.\cite{Battaglia:03}
Setting $\rho_\odot=0.3$ GeV cm$^{-3}$ and $\gamma_{sp}=2.4$,
and assuming that the dark matter cusp follows $\rho\propto r^{-1.5}$
outside of the spike and inside the Solar circle
(the ``adiabatically compressed'' version of the SHM, one finds 
$1.2\times 10^4\lesssim \Delta\Omega\overline{J} \lesssim 1.5\times 10^6$.
In the absence of the spike, 
$\overline{J}$ values are many orders of magnitude lower.
Such a wide range of $\overline{J}$ values makes it difficult
to constrain $m$ or $\langle\sigma v\rangle$ from a
measured annihilation flux or even to conclude that a signal
would be detectable.

Fortunately, many of the models in Fig. 1 can be ruled out.
Once the dark matter distribution has been set up, 
it will evolve, due to interactions between dark matter
particles, stars, and the SBH. 
In most (though not all) circumstances, this evolution has
the effect of decreasing the dark matter density at $r\lesssim r_h$.
The most important evolutionary mechanisms are:

\begin{itemize}

\item {\it Scattering of dark matter particles off of stars.}
Stars in the Galactic nucleus have much larger kinetic energies
than dark matter particles, and gravitational encounters between
the two populations tend to drive them toward mutual equipartition,
$(1/2)m_\star v_\star^2\approx (1/2) m v^2$.
Since the stars are the dominant population (at least at the 
current epoch), the dark matter
heats up, in a time \cite{Merritt:04b}
\be
T_{\rm heat} \approx 10^9 {\rm yr}\times 
\left({M_{bh}\over 3\times 10^6M_\odot}\right)^{1/2} 
\left({r_h\over 2\ {\rm pc}}\right)^{3/2} 
\left({\tilde m_\star\over M_\odot}\right)^{-1}
\label{eq:theat}
\ee
with $m_\star$ the mean stellar mass.
This heating tends to lower the dark matter density and
leads ultimately (in a time of several $ T_{\rm heat}$)
to a density profile of the form $\rho\sim r^{-3/2}$, 
$r_{\rm Sch}\lesssim r \lesssim r_h$,
i.e. it flattens a pre-existing spike.\cite{Merritt:04b,Gnedin:04,Ilyin:04} 
$T_{\rm heat}$ is also roughly the time for exchange
of kinetic energy {\it between} stars, as a result of which
the {\it stellar} density profile itself evolves toward a steady-state
form, though with a steeper index
$\rho_\star\sim r^{-7/4}$.\cite{Bahcall:76,Preto:04} 
This is probably the origin of the power-law 
stellar density cusp at the Galactic center (eq. \ref{eq:nucdens}); 
the observed index, $-1.4$, is not quite as steep as the theoretical value
but is probably consistent given the measurement uncertainties
and given that the Galactic center contains stars with a range
of masses and luminosities.
\smallskip

Interestingly, even if the stellar and dark matter cusps
were once destroyed by the scouring effect of a binary 
SBH,\cite{Milos:01,Ullio:01,Merritt:02}, 
both might have been re-generated by this mechanism.
In the case of the stars, as long as the density within
the SBH's influence radius remains large enough that the
star-star relaxation time is shorter than $\sim 10^{10}$ yr,
the $r^{-7/4}$ cusp will re-form via the Bahcall-Wolf mechanism
\cite{Bahcall:76}.
This is likely to be the case if the mass ratio of the
pre-existing binary SBH was extreme, e.g. $\sim 10:1$.
Once the stellar cusp is back in place, heating of dark matter
particles by stars will drive the dark matter towards its
steady-state distribution, $\rho\sim r^{-3/2}$, although
with a lower normalization than before cusp destruction.
While many such evolutionary scenarios are possible
(and should be worked out in more detail),
the existence of a dense, collisional
nucleus of stars like that observed in the Galactic center
is strong circumstantial evidence of a steeply-rising 
{\it dark} matter density near the SBH.

\item{\it Capture of dark matter within the SBH.}
Any dark matter particles on orbits that intersect the 
SBH are lost in a single orbital period.
Subsequently, scattering of dark matter particles off 
stars drives a continuous flux of dark matter into 
the SBH.\cite{Berezinsky:94b}
Changes in orbital angular momentum dominate the flux;
in a time $T_{\rm heat}$, most of the dark matter within
$r_h$ will have been lost, although the net change in the 
dark matter density profile will be more modest than this
suggests since more particles are continuously being scattered in.\cite{Merritt:04b}

\item {\it Capture of dark matter within stars.}
Another potential loss term for the dark matter is 
capture {\it within} stars, due to scattering off nuclei
followed by annihilation in stellar cores.\cite{Press:85,Salati:89,Bouquet:89}
However this process is not likely to be important
unless the cross section for WIMP-on-proton scattering is very large.

\end{itemize}

These effects, as well as dark matter self-annihilations,
can be modelled in a time-dependent way via the orbit-averaged
Fokker-Planck equation:\cite{Merritt:04b}
\begin{equation}
{\partial f\over\partial t} = -{1\over 4\pi^2p} 
{\partial F_E\over\partial E} - 
f(E)\nu_{\rm coll}(E) - f(E)\nu_{\rm lc}(E)\\
\label{eq:fp}
\end{equation}
with $f(E)$ the phase-space mass density of dark matter,
$E\equiv -v^2/2+\phi(r)$ the energy per unit mass of a 
dark-matter particle, and $\phi(r)$ the gravitational
potential generated by the stars and SBH.
The terms on the right-hand side of Eq. (\ref{eq:fp})
describe the diffusion of dark matter particles in phase space
space due to scattering off stars ($F_E$); 
loss of dark matter due to self-annihilations and/or 
capture within stars ($\nu{\rm coll}$);
and the loss-cone flux into the SBH ($\nu_{\rm lc}$.
Eq. (\ref{eq:fp}) assumes an isotropic distribution of
dark-matter velocities; this assumption is likely to break
down very near the SBH, but the angular-momentum dependence
of  the loss cone flux is well understood and can be
approximated via an energy-dependent loss term 
$\nu_{\rm lc}(E)$.\cite{Merritt:04b}

\begin{figure}[th]
\centerline{\psfig{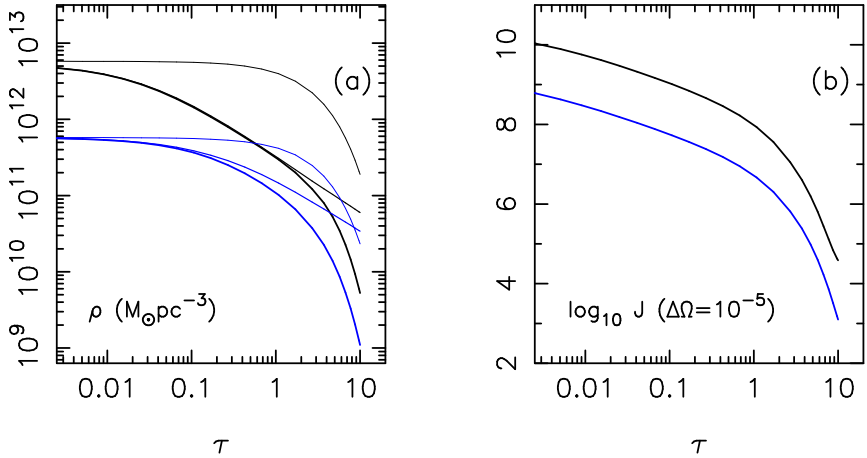}}
\vspace*{6pt}
\caption{(a) Evolution of the dark matter density at a radius
of $10^{-5}r_h\approx 2\times 10^{-5}{\rm pc}$ from the 
center of the Milky Way due to the
physical processes discussed in the text (Eq. \ref{eq:fp}).
Initial conditions were the standard halo model discussed in the text
plus spike; the upper(lower) set of 
curves correspond to an initial density normalization
at $r_h$ of $10(100)M_\odot$ pc$^{-3}$.
Mass and annihilation cross section of the dark
matter particles were set to $m=200$ GeV, 
$\langle\sigma v\rangle=10^{-27}{\rm cm}^3{\rm s}^{-1}$.
In order of increasing thickness, the curves show the
evolution of $\rho$ in response to heating by stars;
to self-annihilations; and to both processes acting
together.
Time is in units of $T_{\rm heat}$ defined in the text;
$\tau=10$ corresponds roughly to $10^{10}$ yr.
(b) Evolution of $\overline{J}$ averaged over an angular
window of $10^{-5}$ sr.
}
\end{figure}

Fig. 2a shows the evolution of the dark matter density
computed in this way, starting from a $\rho\sim r^{-2.33}$
spike ($\rho\sim r^{-1}$ cusp).
Two values were taken for the initial density normalization
at $r=r_h$, $\rho(r_h)=(10,100)M_\odot{\rm pc}^{-3}$;
these values bracket the value $\rho(r_h)\approx 30M_\odot{\rm pc}^{-3}$
obtained by extrapolating the SHM
inwards from the Solar circle with $\rho(R_\odot)=0.3 {\rm GeV} {\rm cm}^{-3}$
(Fig. 1).
The self-annihilation term in Eq. (\ref{eq:fp}) was computed
assuming $m=200$ GeV, 
$\langle\sigma v\rangle=10^{-27}{\rm cm}^3{\rm s}^{-1}$.
The early evolution is dominated by self-annihilations
but for $t\gtrsim 10^9{\rm yr}\approx T_{\rm heat}$,
heating of dark matter by stars dominates.
The change in $\overline{J}(\Delta\Omega=10^{-5})$ (Fig. 2b.) is dramatic,
with final values in the range $10^4\lesssim\overline{J}\lesssim 10^5$.

\section {Direct and Indirect Detection}

In order to understand the nature of the dark matter, 
it is crucial to search for non-gravitational signatures.
Physics beyond the standard model is actively being investigated
using accelerators, and this is one of the main goals of the
upcoming Large Hadron Collider (LHC), expected to begin operation around
2007 with proton-proton collisions at center-of-mass energies
of 14 TeV. 
Numerous classes of models which provide interesting dark matter 
candidates will be tested at the LHC 
(Refs. \refcite{atlas} -- \refcite{lhcreach7}).

An alternative approach is provided by so-called {\it direct detection} 
experiments. 
If the Galaxy is filled with WIMPs, then many of them should 
pass through the Earth, making it possible to look for the interaction of such 
particles with baryonic matter, 
e.g. by recording the recoil energy of nuclei as WIMPs 
scatter off them.\cite{direarly1,direarly2} 
The signal in this case depends on the density and velocity 
distribution of WIMPs in the solar neighborhood and on the WIMP-nucleon 
scattering cross section. 
WIMPs can couple either with the spin content of a nucleus, through
axial-vector (spin dependent) interactions, or with the nuclear mass, 
through scalar  (spin-independent) scattering. The second interaction 
typically dominates over spin-dependent scattering in current experiments, 
which use heavy atoms as targets.
More than 20 direct dark matter detection experiments are either now operating 
or are currently in development. 
Presently, the best direct detection limits come from the 
CDMS~\cite{Akerib:2003px} and Edelweiss~\cite{Benoit:2002hf}
experiments, which probe nucleon-WIMP scattering cross sections of 
order $10^{-7}$pb, for a WIMP mass of order 100 GeV. 
These experiments have ruled out the WIMP discovery claimed by the DAMA 
collaboration,\cite{Bernabei:2003xg} 
although it may still be possible to find 
exotic particle candidates and Galaxy halo models which are able to 
accommodate the data from all current 
experiments.\cite{Prezeau:2003sv,Gelmini:2004gm,Tucker-Smith:2004jv}

Alternatively, dark matter can be searched for {\it indirectly}, 
via the study of the local flux of positrons and antiprotons. 
In fact, if the paradigm of WIMPs as massive particles in thermal 
equilibrium in the early universe is correct, dark matter
particles are expected to annihilate in the Galactic halo producing possibly
large fluxes of secondary particles. 
The High Energy Antimatter Telescope (HEAT) observed a flux of cosmic 
positrons well in excess of the predicted rate, 
peaking around $\sim10 {\rm GeV}$ and extending to higher 
energies,\cite{heat1995} which could be the product of dark matter 
annihilations if sufficient clumping were present in the halo 
(Refs. \refcite{positrons1} -- \refcite{Cheng:2002ej}).
Upcoming experiments, such as AMS-02,\cite{ams02} PAMELA,\cite{pamela} and 
Bess Polar,\cite{besspolar} 
will refine the positron spectrum considerably. 
An antiproton signal could also provide a signature of 
dark matter.\cite{antiproton1,Donato:2003xg,Bergstrom:1999jc}

Unlike other secondary particles, neutrinos produced by dark matter
annihilations can escape from dense media in which such annihilations may take
place. 
For example, WIMPs that are captured in the Sun or Earth can
annihilate at great rates. 
Although gamma rays cannot escape these objects, 
neutrinos often can, providing an interesting signature to search for with 
high-energy neutrino telescopes (Refs. \refcite{indirectneutrino1} -- 
\refcite{indirectneutrino8}).
Although strongly model-dependent, the limits on the neutrino flux can be 
used to set constraints on dark matter particles. 

Dark matter could also be indirectly detected through its 
annihilation radiation in the Galactic halo or in extra-galactic sources. 
The prospects of detecting synchrotron radiation due to dark 
matter annihilations in the Galactic center have been studied in
Refs. ~\refcite{Gondolo:2000pn} --\refcite{Aloisio:2004hy}, while
Ref.~\refcite{Bertone:2004ag} contains a discussion of the prospects for
detecting the neutrino flux. 
Annihilation radiation could be enhanced by the presence of substructures
in the Galactic halo, either ``clumps'' 
(Refs. \refcite{Bergstrom:1998zs} -- \refcite{Koushiappas:2003bn})
or ``caustics'' (Refs.~\refcite{Sikivie:1999jv} -- ~\refcite{Mohayaee:2005fj}).
Recently, high-resolution numerical simulations have pointed to
the existence of mini-halos with masses as small as $10^-6$ solar
masses and sizes as small as the solar system \cite{Diemand:2005hw} (see 
also Refs. \refcite{green05,loeb05}). If confirmed, this could
have important consequences for indirect DM searches, and certainly
deserves further investigation.

The gamma-ray extra-galactic background produced by dark matter
annihilations taking place in all structures and substructures in 
the Universe has been investigated in Refs. ~\refcite{Bergstrom:2001jj} --
~\refcite{Ullio:2002pj}. One usually
finds that the prospects to observe this signal are less promising
than in the case of the gamma-rays from the Galactic center however
(see e.g. Ref.~\refcite{Ando:2005hr}).

Much attention has been devoted to the study of gamma-rays from dark
matter annihilations in the Galactic center (Refs. \refcite{Stecker:88} 
-- \refcite{Fornengo:2004kj}). 
Given a detector with angular acceptance $\Delta\Omega$ sr,
the observed flux of photons produced by
annihilations of dark matter particles of mass $m$ 
and density $\rho$ is\cite{Bergstrom:97}
\begin{equation}
\Phi(\Delta\Omega,E)=\Delta\Omega{dN\over dE} {\langle\sigma v\rangle\over 4\pi m^2}
\overline{J}_{\Delta\Omega}
\label{eq:j}
\end{equation}
where $dN/dE$ is the spectrum of secondary photons per annihilation,
$\langle\sigma v\rangle$ is the velocity-averaged self-annihilation 
cross section, $dl$ is an element of length along the line of sight and
$\overline{J}_{\Delta\Omega}$ was defined in Eq. (\ref{eq:jdo}).
The flux from solid angle $\Delta\Omega$ is then
\begin{equation}
\Phi(\Delta\Omega,E)\approx 1.9\times 10^{-12}{dN\over dE} 
{\langle\sigma v\rangle\over 10^{-26}{\rm cm}^{-3} {\rm s}^{-1}} 
\left({1{\rm TeV}\over m}\right)^2
\overline{J}_{\Delta\Omega}\Delta\Omega{\rm cm}^{-2} {\rm s}^{-1}.
\end{equation}
Inward extrapolation of the density of the standard halo model (SHM)
from the Solar circle into the Galactic center
gives $\overline{J}_3\approx 10^2$ and $\overline{J}_5\approx 10^3$.
These $\overline{J}$-values are large enough to produce
observable signals for many interesting choices
of $\langle\sigma v\rangle$ and $m$.
On the other hand, a recent detection\cite{Aharonian:2004wa} 
 by ACTs of a Galactic center gamma-ray source with energies up
to 10 TeV requires $\overline{J}$ values that
are $10^2$ to $10^4$ times bigger than in the 
SHM,\cite{Bergstrom:2004cy,Hooper:2004fh}
assuming that the signal comes from dark matter annihilations.
Achieving such large $\overline{J}$ values without abandoning the
$\Lambda$CDM paradigm would require a significant enhancement in the
dark matter density very near the Galactic center, for instance,
in the form of a density ``spike'' near the SBH.

As discussed in \S 3, a steeply-rising dark matter density will
due to self-annihilations and to dynamical interactions with the baryons.
The results from a large set of time integrations of Eq. (\ref{eq:fp})
are summarized in Table 1.
Two extreme particle physics models were considered.
In the first case, in order to maximize the ratio 
$\langle\sigma v\rangle/m$,
a cross section 
$\langle\sigma v\rangle_{\rm{th}} = 3\times 10^{-26}$ cm$^3$ s$^{-1}$
was assumed and $m$ was set to $50$ GeV. 
Higher values of $\langle\sigma v\rangle$, though possible, would imply a low
relic density, making the candidate a subdominant component of the
dark matter in the universe.
The lower limit on the mass strictly applies only to neutralinos in 
theories with gaugino and sfermion mass unification at the GUT 
scale,\cite{Eidelman:2004wy} while the limit on the mass of KK particles 
is higher. 
The second extreme case assumed $\langle\sigma v\rangle=0$.
Table 1 gives values of $\overline{J}(\Delta\Omega=10^{-3})$ 
and $\overline{J}(\Delta\Omega=10^{-5})$ at $\tau=10$ for 
each of these extreme particle physics models and for a variety
of initial condtions.
The final $\overline{J}$-values depend appreciably on the 
particle physics model
only when the initial dark matter density has a spike
around the SBH; in other cases the central density is too low
for annihilations to affect $\overline{J}$.
Particularly in the case of maximal $\langle\sigma v\rangle$,
the final $\overline{J}$ values are modest, 
$\log_{10}\overline{J}_3\lesssim 5.3$ and 
$\log_{10}\overline{J}_5\lesssim 7.0$,
compared with the much larger values at $\tau=0$ in the presence
of spikes.

\begin{table}[h]
\tbl{Results of time-integrations of Eq. (\ref{eq:fp})
from a variety of initial dark matter models.
$\gamma_c$ and $\gamma_{sp}$ are the power-law indices
of the initial cusp and spike respectively.
$r_c$ is the dark matter core radius in units of $r_h\approx 2$ pc; 
$\rho\propto r^{-1/2}$ for $r_{sp}<r<r_c$.
Density at $R_\odot$ is in units of GeV cm$^{-3}$.
$\overline{J}_3$ and $\overline{J}_5$ are
values of $\overline{J}$ averaged over windows of solid angle
$10^{-3}$ sr and $10^{-5}$ sr respectively and normalized
as described in the text.
The final two columns give $\overline{J}$ in evolved models
for $\langle\sigma v\rangle=0$ (no annihilations)
and for $\left(\langle\sigma v\rangle, M\right)=
\left(3\times 10^{-26}\mathrm{cm}^3\mathrm{s}^{-1}, 50 \mathrm{GeV}\right)$
(maximal annihilation rate), respectively.
}
{\begin{tabular}{cccccccc} \toprule
& & & & & & $\log_{10} \overline{J}_3$ ($\overline{J}_5$) &\\
Model & $\gamma_c$ & $\gamma_{sp}$ & $r_c$ & $\rho(R_\odot)$ & 
$\tau=0$ & $\tau=10$ & $\tau=10$ \\
1 & 1.0 & --   & -- & 0.3 & 2.56(3.51) & 2.56(3.50) & 2.56(3.50) \\
2 & 1.0 & --   & -- & 0.5 & 3.00(3.96) & 3.00(3.94) & 3.00(3.94) \\
3 & 1.0 & --   & 10 & 0.3 & 2.54(3.33) & 2.54(3.33) & 2.54(3.33) \\
4 & 1.0 & --   &100 & 0.3 & 2.38(2.65) & 2.38(2.65) & 2.38(2.65) \\
5 & 1.0 & 2.33 & -- & 0.3 & 9.21(11.2) & 3.86(5.84) & 2.56(3.52) \\
6 & 1.0 & 2.33 & -- & 0.5 & 9.65(11.7) & 4.31(6.29) & 3.00(3.96) \\
7 & 1.0 & 2.29 & 10 & 0.3 & 6.98(8.98) & 2.61(3.88) & 2.54(3.33) \\
8 & 1.0 & 2.29 &100 & 0.3 & 5.98(7.98) & 2.39(2.99) & 2.38(2.65) \\
9 & 1.5 & --   & -- & 0.3 & 5.36(7.30) & 4.81(6.58) & 4.78(6.53) \\
10 & 1.5 & --   & -- & 0.5 & 5.80(7.75) & 5.26(7.03) & 5.23(6.98) \\
11 & 1.5 & --   & 10 & 0.3 & 4.51(5.82) & 4.51(5.82) & 4.51(5.82) \\
12 & 1.5 & --   &100 & 0.3 & 3.85(4.23) & 3.85(4.23) & 3.85(4.23) \\
13 & 1.5 & 2.40 & -- & 0.3 & 14.3(16.3) & 8.81(10.8) & 4.81(6.58) \\
14 & 1.5 & 2.40 & -- & 0.5 & 14.8(16.8) & 9.25(11.3) & 5.25(7.02) \\
15 & 1.5 & 2.29 & 10 & 0.3 & 9.67(11.7) & 4.77(6.51) & 4.51(5.82) \\
16 & 1.5 & 2.29 &100 & 0.3 & 7.67(9.67) & 3.87(4.64) & 3.86(4.23) \\
\end{tabular}}
\end{table}

In these evolutionary models, the value of $\overline{J}$ at 
a fixed time $t=\tau T_{\rm heat}$ is determined completely
by the initial conditions and by the quantity 
$\langle\sigma v\rangle/m$ that specifies the annihilation rate.
This outcome can be expressed in terms of the {\it boost factor}
$b$ defined as $\overline{J}/\overline{J}_N$, with $\overline{J}_N$
the value in the SHM having the same
density normalization at $r=R_\odot$ as in the evolving model.
One finds\cite{BM:05} that the boost factors at $\tau=10$
(roughly $10^{10}$ yr) can be well approximated by the function
\begin{equation}
B(X)=B_{max}-(1/2)(B_{max}-B_{min})\{1+\tanh[a(X-b)]\}
\end{equation}
with 
$X\equiv\log_{10}(\langle\sigma v\rangle/10^{-30}{\rm cm}^3{\rm s}^{-1})/
(m/100{\rm GeV})$
and $B\equiv\log_{10}b$.
Values of the fitting parameters are given in Table 2 for the
``spike'' models of Table 1; as noted above, in the absence
of a spike, the final $\overline{J}$ values are essentially
unaffected by annihilations hence independent of 
$\langle\sigma v\rangle/m$.
Recent analyses\cite{Bergstrom:2004cy,Hooper:2004fh}
 of the HESS Galactic center data\cite{Aharonian:2004wa}
suggest that the observed gamma-ray spectrum
is consistent with particle masses of order 10 TeV
and cross sections of order $3\times 10^{-26}{\rm cm}^3{\rm s}^{-1}$,
if boost factors are as high as $10^3\lesssim b\lesssim 10^4$.
Table 2 shows that such boost factors are indeed achievable 
at $X=\log_{10}(3\times 10^{-26}/10^{-30})/(10^4/100)\approx 2.5$
if the initial dark matter distribution is similar to that of 
Models 13 or 14, i.e. a ``spike'' inside of a $\rho\sim r^{-1.5}$ cusp.

\begin{table}
\tbl{Parameters in the fitting function for the boost.}
{\begin{tabular}{ccccccccccc} \toprule
&  & & $\Delta\Omega=$ $10^{-3}$ & & & & & $\Delta\Omega=10^{-5}$ & \\
 Model & & $B_{min}$ & $B_{max}$ & $a$ & $b$ & & $B_{min}$ & $B_{max}$ & $a$ & $b$\\
 5 & & -0.02 & 1.31 & 0.66 & 0.73 && -0.05 & 2.35 & 0.55 & 1.50 \\
 6 & & -0.01 & 1.31 & 0.66 & 0.51 && -0.06 & 2.34 & 0.56 & 1.28 \\
 7 & & -0.02 & 0.05 & 0.75 & 0.92 && -0.18 & 0.38 & 0.72 & 1.31 \\
 8 & & -0.18 & -0.17& 0.76 & 1.37 && -0.86 & -0.51& 0.73 & 1.64 \\
13 & & 2.14  & 6.28 & 0.42 & -0.05&& 2.95  & 7.35 & 0.40 & 0.10 \\  
14 & & 2.16  & 6.29 & 0.43 & -0.28&& 2.97  & 7.36 & 0.41 & 0.13 \\
15 & & 1.96  & 2.21 & 0.74 & -0.27&& 2.32  & 3.02 & 0.71 & 0.07 \\
16 & & 1.30  & 1.31 & 0.75 & 0.53 && 0.73  & 1.14 & 0.73 & 0.84 \\ 
\end{tabular}}
\end{table}

Annihilation radiation might also be detected from the centers
of galaxies other than the Milky Way
(Refs. \refcite{Gondolo:94} -- \refcite{Evans:04}).
Even globular clusters\cite{Giraud:03} have been suggested as
possible targets.
As in the case of the Milky Way, the major uncertainty in 
predictions of the annihilation flux from external galaxies
is the unknown distribution of dark matter on small scales;
absent power-law cusps or spikes, indirect detection of dark
matter from external galaxies is probably impossible with 
the current generation of detectors.\cite{Evans:04}
As argued above, a plausible guide to the dark matter distribution
on small scales is the {\it stellar} density profile.
The Local Group galaxies M31 and M32 both exhibit 
steep stellar cusps, $\rho_\star\sim r^{-\gamma_\star}, 
1.5\lesssim\gamma_\star\lesssim 2.0$, inward of $\sim 1$ pc,
similar to what is observed in the Milky Way;\cite{Lauer:98}
these galaxies also show clear kinematical evidence for SBHs.
By contrast, the Local Group galaxies M33 and NGC205 exhibit
cores in the stellar density and any SBHs in these
galaxies are too small to be detected.\cite{Merritt:01,Valluri:05}
Stellar cusps like the one in the Milky Way would be
unresolved at the distance of the Virgo cluster, but
it is a reasonable guess that all comparably-luminous
galaxies contain similar distributions of luminous
and dark matter on sub-parsec scales.
The situation is likely to be different for giant galaxies like 
M87 in Virgo, which are known to have much shallower stellar density
profiles within their SBH's sphere of influence,
probably a consequence of an earlier epoch of scouring
by binary SBHs.\cite{Milos:02}
The matter distribution near the centers 
of giant galaxies is probably essentially unchanged since 
the last round of galaxy mergers, making them unfelicitous
sites for indirect detection  of dark matter.


\section*{Acknowledgments}
GB is supported by the DOE and the NASA grant NAG 5-10842 
at Fermilab.
DM acknowledges support from the NSF
(AST-0206031, AST-0420920 and AST-0437519),
NASA (NNG04GJ48G)
and the Space Telescope Science Institute 
(HST-AR-09519.01-A and HST-GO-09401.10A).


\begin{thebibliography}{0}


\bibitem{Spergel:2003cb}
D.~N.~Spergel {\it et al.},
Astrophys.\ J.\ Suppl.\  {\bf 148}, 175 (2003)
[arXiv:astro-ph/0302209].

\bibitem{Olive:2003iq}
K.~A.~Olive,
TASI lectures on dark matter,
arXiv:astro-ph/0301505.

\bibitem{Bertone:2004pz}
G.~Bertone, D.~Hooper and J.~Silk,
Phys.\ Rept.\  {\bf 405} (2005) 279
[arXiv:hep-ph/0404175].

\bibitem{Bergstrom:2000pn}
  L.~Bergstrom,
  Rept.\ Prog.\ Phys.\  {\bf 63} (2000) 793
  [arXiv:hep-ph/0002126].

\bibitem{Appelquist:2000nn}
  T.~Appelquist, H.~C.~Cheng and B.~A.~Dobrescu,
  Phys.\ Rev.\ D {\bf 64}, 035002 (2001)
  [arXiv:hep-ph/0012100].

\bibitem{Cheng:2002iz}
H.~C.~Cheng, K.~T.~Matchev and M.~Schmaltz,
Phys.\ Rev.\ D {\bf 66}, 036005 (2002)
[arXiv:hep-ph/0204342].

\bibitem{Servant:2002aq}
G.~Servant and T.~M.~Tait,
Nucl.\ Phys.\ B {\bf 650} (2003) 391
[arXiv:hep-ph/0206071].


\bibitem{leeweinberg}
B.~W.~Lee and S.~Weinberg,
Phys.\ Rev.\ Lett.\  {\bf 39}, 165 (1977).

\bibitem{Hut:1977zn}
P.~Hut,
Phys.\ Lett.\ B {\bf 69} (1977) 85.

\bibitem{weinberg} 
S. Weinberg, 1995, The Quantum Theory of Fields 
Vol 1: Foundations, Cambridge University Press.

\bibitem{Griest:1989wd}
K.~Griest and M.~Kamionkowski,
Phys.\ Rev.\ Lett.\  {\bf 64} (1990) 615.


\bibitem{mevproposed1}
C.~Boehm, T.~A.~Ensslin and J.~Silk,
arXiv:astro-ph/0208458.


\bibitem{mevproposed2}
C.~Boehm and P.~Fayet,
arXiv:hep-ph/0305261.

\bibitem{Jean:2003ci}
  P.~Jean {\it et al.},
   ``Early SPI/INTEGRAL measurements of galactic 511 keV line emission from
   positron annihilation,''
  %
  Astron.\ Astrophys.\  {\bf 407}, L55 (2003)
  [arXiv:astro-ph/0309484].

\bibitem{Knodlseder:2003sv}
  J.~Knodlseder {\it et al.},
   ``Early SPI/INTEGRAL contraints on the morphology of the 511 keV line
   emission in the 4th galactic quadrant,''
  %
  arXiv:astro-ph/0309442.

\bibitem{Weidenspointner:2004my}
  G.~Weidenspointner {\it et al.},
   ``SPI observations of positron annihilation radiation from the 4th galactic
   quadrant: sky distribution,''
  %
  arXiv:astro-ph/0406178.

\bibitem{Beacom:2004pe}
  J.~F.~Beacom, N.~F.~Bell and G.~Bertone,
  arXiv:astro-ph/0409403.

\bibitem{Boehm:2004gt}
  C.~Boehm and Y.~Ascasibar,
  Phys.\ Rev.\ D {\bf 70} (2004) 115013
  [arXiv:hep-ph/0408213].

\bibitem{Boehm:2004uq}
  C.~Boehm,
  Phys.\ Rev.\ D {\bf 70} (2004) 055007
  [arXiv:hep-ph/0405240].

\bibitem{Chung:2001cb}
  D.~J.~H.~Chung, P.~Crotty, E.~W.~Kolb and A.~Riotto,
  Phys.\ Rev.\ D {\bf 64} (2001) 043503
  [arXiv:hep-ph/0104100].

\bibitem{Chang:1996vw}
S.~Chang, C.~Coriano and A.~E.~Faraggi,
Nucl.\ Phys.\ B {\bf 477}, 65 (1996)
[arXiv:hep-ph/9605325].

\bibitem{Primack:03}
  J. R. Primack,
  in {\it Proceedings of the Fifth International UCLA Symposium
          on Sources and Detection of Dark Matter}, 
  Nucl. Phys. B. Proc. Suppl. {\bf 124}, 3
  (astro-ph/0205391)

\bibitem{Merrifield:04}
  M. Merrifield,
  in {\it International Astronomical Union Symposium no. 220}, 
    eds. S. D. Ryder {\it et al.}
    (Astronomical Society of the Pacific, 2004), p. 431

\bibitem{Navarro:04}
  J. N. Navarro,
  in {\it International Astronomical Union Symposium no. 220}, 
    eds. S. D. Ryder {\it et al.}
    (Astronomical Society of the Pacific, 2004), p. 61

\bibitem{Navarro:04b}
	J. F. Navarro {\it et al.},
	{\it Mon. Not. R. Astron. Soc.} {\bf 349}, 1039 (2004).

\bibitem{Reed:05}
  D.~Reed, F.~Governato, L.~Verde, J.~Gardner, T.~Quinn, J.~Stadel, J., 
  D.~Merritt, G.~Lake,
  {\it Mon. Not. R. Astron. Soc.} {\bf 357}, 82 (2005)

\bibitem{Merritt:05}
  D.~Merritt, J.~Navarro, A.~Ludlow, and A.~Jenkins, 
  astro-ph/0502515 (2005)

\bibitem{NFW:96}
        J.~F. Navarro, C.~S. Frenk, and S.~D.~M.~White,
	{\it Astrophys. J.} {\bf 462}, 563 (1996).

\bibitem{Blumenthal:86}
        G.~R.~Blumenthal, S.~M.~Faber, R.~Flores, and J.~R.~Primack,
	Astrophys.\ J. {\bf 301}, 27 (1986).

\bibitem{Edsjo:04}
  J.~Edsjo, M.~Schelke, and P.~Ullio,
  astro-ph/0405414.

\bibitem{Prada:04}
  F.~Prada, A.~Klypin, J.~Flix, M.~Martinez, and E.~Simonneau,
  Phys.\ Rev.\ Lett. {\bf 93}, 241301 (2004).

\bibitem{Gnedin:04a}
  O.~Y.~Gnedin, A.~V.~Kravtsov, A.~A.~Klypin, and D.~Nagai,
  Astrophys.~J. {\bf 616}, 16 (2004).

\bibitem{Ullio:01}
	P. Ullio, H.-Z. Zhao, and M. Kamionkowski,
	{\it Phys. Rev. D.}, {\bf 64}, 043504 (2001).

\bibitem{Merritt:02}
	D. Merritt, M. Milosavljevic, L. Verde, and R. Jimenez,
	Phys. Rev. Lett. {\bf 88}, 191301 (2002).

\bibitem{MM:02}
  D. Merritt and M. Milosavljevic,
  in {\it Dark Matter in Astro- and Particle Physics}, 
  eds. H. V. Klapdor-Kleingrothaus and R. D. Viollier
  (Springer, 2002), p. 79.

\bibitem{Genzel:03}
	R.~Genzel et al.,
	{\it Astrophys. J.} {\bf 594}, 812 
	(2003).

\bibitem{Chakrabarty:01}
  D.~Chakrabarty and P.~Saha,
  {\it Astronom. J.} {\bf 122}, 232 
  (2001).

\bibitem{Ghez:03}
        A. M. Ghez et al.
	{\it Astrophys. J.} {\bf 586}, L127
	(2003).

\bibitem{Schoedel:03}
	R. Schoedel, R. Genzel, T. Ott, A. Eckart, N. Mouawad
	and T. Alexander,
	{\it Astrophys. J.} {\bf 596}, 1015
	(2003).

\bibitem{Peebles:72}
  P. J. E. Peebles,
  {\it Gen. Rel. Grav.} {\bf 3}, 61 (1972).

\bibitem{Young:80}
  P. J. Young,
  {\it Astrophys. J.} {\bf 242}, 1232 (1980)

\bibitem{Gondolo:99}
	P. Gondolo and J. Silk,
	Phys. Rev. Lett. {\bf 83}, 1719 (1999).

\bibitem{Merritt:04a}
	D.~Merritt, 
	in Carnegie Observatories Astrophysics Series, Vol. 1: 
	Coevolution of Black Holes and Galaxies, ed. L. C. Ho 
	(Cambridge: Cambridge Univ. Press), in press
	(2004).

\bibitem{Berezinsky:94}
  V. Berezinsky, A. Bottino, and G. Mignola,
  {\it Phys. Lett. B} {\bf 325}, 136 (1994).

\bibitem{EGRET}
	http://cossc.gsfc.nasa.gov/egret/

\bibitem{VERITAS}
	http://veritas.sao.arizona.edu/

\bibitem{CANGAROO}
  http://icrhp9.icrr.u-tokyo.ac.jp/

\bibitem{HESS}
        http://www.mpi-hd.mpg.de/hfm/HESS/HESS.html/

\bibitem{GLAST}
	http://www-glast.stanford.edu/


\bibitem{Battaglia:03}
  M.~Battaglia, A.~De Roeck, J.~R.~Ellis, F.~Gianotti, K.~A.~Olive,
  and L.~Pape,
  {\it Eur.\ Phys.\ J.\ C} {\bf 33}, 273 (2004).

\bibitem{Merritt:04b}
  D.~Merritt,
  Phys.\ Rev.\ Lett.\  {\bf 92}, 201304 
  (2004).

\bibitem{Ilyin:04}
  A. S. Ilyin, K. P. Zybin, and A. V. Gurevich,
   Zh. Eksp. Teor. Fiz. {\bf 98}, 5 
   (2004).

\bibitem{Gnedin:04}
	O. Y. Gnedin and J. R. Primack,
	Phys. Rev. Lett. {\bf 93}, 061302
	(2004).

\bibitem{Bahcall:76}
  J. N. Bahcall and S. Wolf,
  Astrophys. J. {\bf 209}, 214
  (1976).

\bibitem{Preto:04}
  M. Preto, D. Merritt, and R. Spurzem,
  Astrophys. J. {\bf 613}, L109
  (2004).

\bibitem{Milos:01}
  M. Milosavljevic and D. Merritt,
  Astrophys. J., {\bf 563}, 34 
  (2001).

\bibitem{Berezinsky:94b}
  V. S. Berezinsky, A. V. Gurevich, and K. P. Zybin, 
  {\it Phys. Lett. B} {\bf 294}, 221 (1994).

\bibitem{Press:85}
  W.~H.~Press and D.~N.~Spergel,
  {\it Astrophys. J.}  {\bf 296}, 679 (1985).

\bibitem{Salati:89}
        P.~Salati and J.~Silk,
	{\it Astrophys.\ J.} {\bf 338}, 24 (1989).

\bibitem{Bouquet:89}
  A. Bouquet, P. Salati, and J. Silk,
  {\it Phys. Rev. D} {\bf 40}, 3168 (1989).

\bibitem{atlas}
 ATLAS TDR, report CERN/LHCC/99-15 (1999).


\bibitem{cms}
CMS TP, report CERN/LHCC/94-38 (1994).


\bibitem{Dawson:1983fw}
S.~Dawson, E.~Eichten and C.~Quigg,
Phys.\ Rev.\ D {\bf 31}, 1581 (1985).

\bibitem{Allanach:2004ub}
B.~C.~Allanach {\it et al.}  [Beyond the Standard Model Working Group Collaboration], ``Les Houches 'Physics at TeV Colliders 2003' Beyond the Standard Model
Working Group: Summary report'', arXiv:hep-ph/0402295.


\bibitem{Allanach2}
B.~C.~Allanach, C.~G.~Lester, M.~A.~Parker and B.~R.~Webber,
JHEP {\bf 0009}, 004 (2000)
[arXiv:hep-ph/0007009].

\bibitem{lhcreach1}
H.~Baer, C.~Balazs, A.~Belyaev, T.~Krupovnickas and X.~Tata,
JHEP {\bf 0306}, 054 (2003)
[arXiv:hep-ph/0304303].

\bibitem{lhcreach4}
I.~Hinchliffe, F.~E.~Paige, M.~D.~Shapiro, J.~Soderqvist and W.~Yao,
Phys.\ Rev.\ D {\bf 55}, 5520 (1997)
[arXiv:hep-ph/9610544].

\bibitem{lhcreach6}
G.~Polesello and D.~R.~Tovey,
arXiv:hep-ph/0403047.


\bibitem{lhcreach7}
A.~Birkedal, K.~Matchev and M.~Perelstein,
arXiv:hep-ph/0403004.



\bibitem{direarly1}
A.~Drukier and L.~Stodolsky,
Phys.\ Rev.\ D {\bf 30}, 2295 (1984).

\bibitem{direarly2}
M.~W.~Goodman and E.~Witten,
Phys.\ Rev.\ D {\bf 31}, 3059 (1985).


\bibitem{Akerib:2003px}
D.~S.~Akerib {\it et al.}  [CDMS Collaboration],
Phys.\ Rev.\ D {\bf 68} (2003) 082002
[arXiv:hep-ex/0306001].

\bibitem{Benoit:2002hf}
A.~Benoit {\it et al.},
Phys.\ Lett.\ B {\bf 545} (2002) 43
[arXiv:astro-ph/0206271].

\bibitem{Bernabei:2003xg}
R.~Bernabei {\it et al.},
{\it Talk at the 10th International Workshop on Neutrino Telescopes, Venice, Italy}, (2003), [arXiv:astro-ph/0305542].

\bibitem{Prezeau:2003sv}
G.~Prezeau, A.~Kurylov, M.~Kamionkowski and P.~Vogel,
Phys.\ Rev.\ Lett.\  {\bf 91}, 231301 (2003)
[arXiv:astro-ph/0309115].

\bibitem{Gelmini:2004gm}
G.~Gelmini and P.~Gondolo,
arXiv:hep-ph/0405278.

\bibitem{Tucker-Smith:2004jv}
D.~Tucker-Smith and N.~Weiner,
arXiv:hep-ph/0402065.

\bibitem{heat1995}
S.~W.~Barwick {\it et al.}  [HEAT Collaboration],
Astrophys.\ J.\  {\bf 482}, L191 (1997)
[arXiv:astro-ph/9703192].

\bibitem{positrons1}
G.~L.~Kane, L.~T.~Wang and J.~D.~Wells,
Phys.\ Rev.\ D {\bf 65}, 057701 (2002).

\bibitem{positrons2}
M.~Kamionkowski and M.~S.~Turner,
Phys.\ Rev.\ D {\bf 43}, 1774 (1991).

\bibitem{Baltz:2001ir}
E.~A.~Baltz, J.~Edsjo, K.~Freese and P.~Gondolo,
Phys.\ Rev.\ D {\bf 65} (2002) 063511
[arXiv:astro-ph/0109318].

\bibitem{positrons3}
M.~S.~Turner and F.~Wilczek,
Phys.\ Rev.\ D, {\bf 42}, 1001 (1990).

\bibitem{positrons4}
A.~J.~Tylka,
Phys.\ Rev.\ Lett., {\bf 63}, 840 (1989).

\bibitem{positrons5}
G.~L.~Kane, L.~T.~Wang and T.~T.~Wang,
Phys.\ Lett.\ B {\bf 536}, 263 (2002).

\bibitem{positrons6}
E.~A.~Baltz and J.~Edsjo,
Phys.\ Rev.\ D {\bf 59} (1999) 023511
[arXiv:astro-ph/9808243].

\bibitem{Cheng:2002ej}
H.~C.~Cheng, J.~L.~Feng and K.~T.~Matchev,
Phys.\ Rev.\ Lett.\  {\bf 89}, 211301 (2002)
[arXiv:hep-ph/0207125].

\bibitem{ams02}
R.~Battiston, {\it Frascati Physics Series} {\bf 24}, 261.

\bibitem{pamela}
http://www.cerncourier.com/main/article/42/8/17

\bibitem{besspolar}
T.~Sanuki,
Int.\ J.\ Mod.\ Phys.\ A {\bf 17}, 1635 (2002).

\bibitem{antiproton1}
A.~Bottino, F.~Donato, N.~Fornengo and P.~Salati,
Phys.\ Rev.\ D {\bf 58}, 123503 (1998).


\bibitem{Donato:2003xg}
F.~Donato, N.~Fornengo, D.~Maurin, P.~Salati and R.~Taillet,
arXiv:astro-ph/0306207.

\bibitem{Bergstrom:1999jc}
L.~Bergstrom, J.~Edsjo and P.~Ullio,
arXiv:astro-ph/9902012.


\bibitem{indirectneutrino1}
J.~Silk, K.~Olive and M.~Srednicki,
Phys.\ Rev.\ Lett.\ {\bf 55}, 257 (1985).

\bibitem{indirectneutrino2}
F.~Halzen, T.~Stelzer and M.~Kamionkowski,
Phys.\ Rev.\ D {\bf 45}, 4439 (1992).


\bibitem{indirectneutrino3}
J.~L.~Feng, K.~T.~Matchev and F.~Wilczek,
Phys.\ Rev.\ D {\bf 63}, 045024 (2001).

\bibitem{indirectneutrino4}
V.~D.~Barger, F.~Halzen, D.~Hooper and C.~Kao,
Phys.\ Rev.\ D {\bf 65}, 075022 (2002).

\bibitem{indirectneutrino5}
V.~Berezinsky, A.~Bottino, J.~Ellis, N.~Fornengo, G.~Mignola and S.~Scopel,
Astropart.\ Phys.\ {\bf 5}, 333 (1996)
[hep-ph/9603342].

\bibitem{Bergstrom:1998xh}
L.~Bergstrom, J.~Edsjo and P.~Gondolo,
Phys.\ Rev.\ D {\bf 58} (1998) 103519
[arXiv:hep-ph/9806293].

\bibitem{indirectneutrino6}
K.~Freese,
Phys.\ Lett.\ B {\bf 167}, 295 (1986).

\bibitem{indirectneutrino7}
L.~M.~Krauss, M.~Srednicki and F.~Wilczek,
Phys.\ Rev.\ D {\bf 33}, 2079 (1986).

\bibitem{indirectneutrino8}
T.~K.~Gaisser, G.~Steigman and S.~Tilav,
Phys.\ Rev.\ D {\bf 34}, 2206 (1986).

\bibitem{Gondolo:2000pn}
  P.~Gondolo,
  Phys.\ Lett.\ B {\bf 494}, 181 (2000).

\bibitem{Bertone:2002ms}
  G.~Bertone, G.~Servant and G.~Sigl,
  Phys.\ Rev.\ D {\bf 68} (2003) 044008
  [arXiv:hep-ph/0211342].

\bibitem{Aloisio:2004hy}
  R.~Aloisio, P.~Blasi and A.~V.~Olinto,
  JCAP {\bf 0405}, 007 (2004)
  [arXiv:astro-ph/0402588].

\bibitem{Bertone:2004ag}
  G.~Bertone, E.~Nezri, J.~Orloff and J.~Silk,
  Phys.\ Rev.\ D {\bf 70}, 063503 (2004)
  [arXiv:astro-ph/0403322].

\bibitem{Bergstrom:1998zs}
L.~Bergstrom, J.~Edsjo and P.~Ullio,
Phys.\ Rev.\ D {\bf 58} (1998) 083507
[arXiv:astro-ph/9804050].

\bibitem{Calcaneo-Roldan:2000yt}
C.~Calcaneo-Roldan and B.~Moore,
Phys.\ Rev.\ D {\bf 62} (2000) 123005
[arXiv:astro-ph/0010056].


\bibitem{Tasitsiomi:2002vh}
A.~Tasitsiomi and A.~V.~Olinto,
Phys.\ Rev.\ D {\bf 66} (2002) 083006
[arXiv:astro-ph/0206040].


\bibitem{Berezinsky:2003vn}
V.~Berezinsky, V.~Dokuchaev and Y.~Eroshenko,
Phys.\ Rev.\ D {\bf 68} (2003) 103003
[arXiv:astro-ph/0301551].

\bibitem{Stoehr:2003hf}
F.~Stoehr, S.~D.~White, V.~Springel, G.~Tormen and N.~Yoshida,
Mon.\ Not.\ Roy.\ Astron.\ Soc.\  {\bf 345}, 1313 (2003)
[arXiv:astro-ph/0307026].

\bibitem{Blasi:2002ct}
P.~Blasi, A.~V.~Olinto and C.~Tyler,
Astropart.\ Phys.\  {\bf 18}, 649 (2003)
[arXiv:astro-ph/0202049].

\bibitem{Koushiappas:2003bn}
S.~M.~Koushiappas, A.~R.~Zentner and T.~P.~Walker,
Phys.\ Rev.\ D {\bf 69}, 043501 (2004)
[arXiv:astro-ph/0309464].

\bibitem{Sikivie:1999jv}
P.~Sikivie,
Phys.\ Rev.\ D {\bf 60} (1999) 063501
[arXiv:astro-ph/9902210].

\bibitem{Bergstrom:2000bk}
L.~Bergstrom, J.~Edsjo and C.~Gunnarsson,
Phys.\ Rev.\ D {\bf 63} (2001) 083515
[arXiv:astro-ph/0012346].

\bibitem{Mohayaee:2005fj}
  R.~Mohayaee and S.~Shandarin,
  arXiv:astro-ph/0503163.

\bibitem{Diemand:2005hw}
  J.~Diemand, B.~Moore and J.~Stadel,
  Nature {\bf 433} (2005) 389.


\bibitem{green05}
  A.~M.~Green, S.~Hofmann and D.~J.~Schwarz,
  arXiv:astro-ph/0503387.


\bibitem{loeb05}
  A.~Loeb and M.~Zaldarriaga,
  arXiv:astro-ph/0504112.

\bibitem{Bergstrom:2001jj}
L.~Bergstrom, J.~Edsjo and P.~Ullio,
Phys.\ Rev.\ Lett.\  {\bf 87} (2001) 251301
[arXiv:astro-ph/0105048].

\bibitem{Taylor:2002zd}
J.~E.~Taylor and J.~Silk,
Mon.\ Not.\ Roy.\ Astron.\ Soc.\  {\bf 339} (2003) 505
[arXiv:astro-ph/0207299].

\bibitem{Ullio:2002pj}
P.~Ullio, L.~Bergstrom, J.~Edsjo and C.~Lacey,
Phys.\ Rev.\ D {\bf 66} (2002) 123502
[arXiv:astro-ph/0207125].




\bibitem{Stecker:88}
  F. W. Stecker,
  {\it Phys. Lett. B} {\bf 201}, 529 (1988)

\bibitem{Bergstrom:97}
  L.~Bergstrom, P.~Ullio and J.~H.~Buckley,
  {\it Astropart. Phys.}  {\bf 9}, 137 (1998).

\bibitem{Ando:2005hr}
  S.~Ando,
  arXiv:astro-ph/0503006.




\bibitem{Bertone:2001jv}
  G.~Bertone, G.~Sigl and J.~Silk,
  Mon.\ Not.\ Roy.\ Astron.\ Soc.\  {\bf 326} (2001) 799
  [arXiv:astro-ph/0101134].

\bibitem{Cesarini:2003nr}
  A.~Cesarini, F.~Fucito, A.~Lionetto, A.~Morselli and P.~Ullio,
  Astropart.\ Phys.\  {\bf 21}, 267 (2004)
  [arXiv:astro-ph/0305075].

\bibitem{Fornengo:2004kj}
  N.~Fornengo, L.~Pieri and S.~Scopel,
  Phys.\ Rev.\ D {\bf 70} (2004) 103529
  [arXiv:hep-ph/0407342].

\bibitem{Bergstrom:2004cy}
  L.~Bergstrom, T.~Bringmann, M.~Eriksson, and M.~Gustafsson,
  astro-ph/0410359.

\bibitem{Hooper:2004fh}
  D.~Hooper and J.~March-Russell,
  hep-ph/0412048.

\bibitem{Eidelman:2004wy}
  S.~Eidelman {\it et al.},
  {\it Phys.\ Lett.\ B} {\bf 592}, 1 (2004).

\bibitem{BM:05}
  G. Bertone and D. Merritt,
  astro-ph/0501000

\bibitem{Aharonian:2004wa}
  F.~A.~Aharonian {\it et al.},
  {\it Astron.\ Astrophys.} {\bf 425}, L13 (2004).

\bibitem{Gondolo:94}
  P. Gondolo,
  {\it Nucl. Phys. B (Proc. Suppl)} {\bf 35}, 148 (1994).

\bibitem{Baltz:99}
  E. A. Baltz, C. Briot, P. Salati, R. Taillet, and J. Silk,
  {\it Phys. Rev. D.} {\bf 61}, 023514 (1999).

\bibitem{Tyler:02}
  C. Tyler,
  {\it Phys. Rev. D} {\bf 66}, 023509 (2002).

\bibitem{Aharonian:03}
  F.~A.~Aharonian {\it et al.},
  {\it Astron.\ Astrophys.} {\bf 400}, 153 (2003).

\bibitem{Falvard:03}
  A. Falvard {\it et al.},
  {\it Astropart. Phys.} {\bf 20}, 467 (2003).

\bibitem{Vassiliev:03}
  V. V. Vassiliev {\it et al.},
  in {\it Proceedings of the 28th International Cosmic Ray Conference}
  eds. (Universal Academy Press, 2003).

\bibitem{Pieri:04}
  L. Pieri and E. Branchini,
  {\it Phys. Rev. D} {\bf 69}, 043512 (2004).

\bibitem{Fornengo:04}
  N. Fornengo, L. Pieri, and S. Scopel,
  {\it Phys. Rev. D.} {\bf 70}, 103529 (2004).

\bibitem{Olinto:04}
  A. Tasitsiomi, J. Gaskins, and A. V. Olinto, 
  {\it Astropart. Phys.} {\bf 21}, 637

\bibitem{Hooper:04}
  D. Hooper {\it et al.},
  {\it Phys. Rev. Lett.} {\bf 93}, 161302 (2004).

\bibitem{Evans:04}
  N. W. Evans, F. Ferrer, and S. Sarker,
  {\it Phys. Rev. D} {\bf 69}, 123501 (2004).

\bibitem{Giraud:03}
  E. Giraud {\it et al.},
  in {\it Astronomy, Cosmology and Fundamental Physics, 
    Proceedings of the ESO-CERN-ESA Symposium}, eds. P. A. Shaver, 
  L. Di Lella, and A. Gimenez (Springer, 2003).

\bibitem{Lauer:98}
  T. R. Lauer, S. M. Faber, E. A. Ajhar, C. J. Grillmair, and P. A. Scowen,
  {\it Astron. J.} {\bf 116}, 2263 (1998).

\bibitem{Merritt:01}
  D. Merritt, L. Ferrarese, and C. L. Joseph,
  {\it Science}, {\bf 293}, 1116 (2001).

\bibitem{Valluri:05}
  M. Valluri, L. Ferrarese, D. Merritt, and C. L. Joseph,
  astro-ph/0502493 (2005).

\bibitem{Milos:02}
	M. Milosavljevi\'c, D. Merritt, A. Rest, and F. van den Bosch,
	Mon. Not. R. Astron. Soc. {\bf 331}, L51 (2002).

\end{thebibliography}
\end{document}